\documentclass[preprint, 12pt]{elsarticle}

\usepackage{amssymb, amsmath, amsthm}

\newdefinition{qst}{Question}
\newtheorem{thm}{Theorem}
\newtheorem{lm}{Lemma}
\newdefinition{dfn}{Definition}
\newtheorem{prp}{Proposition}

\newproof{pf}{Proof}
\newproof{pfof}{Proof of Lemma \ref{minterm}}

\DeclareMathOperator{\cnf}{\mathsf{CNF}}
\DeclareMathOperator{\dnf}{\mathsf{DNF}}

\journal{}

\begin{document}

\begin{frontmatter}

\title{On the Structure and the Number of Prime Implicants of 2-$\cnf$s}

\author{Navid Talebanfard \footnote{Part of this work was done while the author was with Aarhus University. He acknowledges support from the Danish
   National Research Foundation and The National Science Foundation
   of China (under the grant 61061130540) for the Sino-Danish Center
   for the Theory of Interactive Computation, and from the CFEM
  research center (supported by the Danish Strategic Research
   Council), within which this work was performed.}}

\ead{navid@is.titech.ac.jp}

\address{Department of Mathematical and Computing Sciences, Tokyo Institute of Technology, Meguro-ku Ookayama 2-12-1, Japan 152-8552}

\begin{abstract}
Let $m(n, k)$ be the maximum number of prime implicants that any $k$-$\cnf$ on $n$ variables can have. We show that $3^{\frac{n}{3}} \le m(n, 2) \le (1 + o(1))3^{\frac{n}{3}}$.
\end{abstract}


\end{frontmatter}

\section{Introduction}

A {\it prime implicant} of a Boolean function $f$ is a maximal subcube contained in $f^{-1}(1)$. It is known that among Boolean functions on $n$ variables the maximum number of prime implicants is between $\Omega(\frac{3^n}{n})$ and $O(\frac{3^n}{\sqrt{n}})$ (see \cite{chand}). It is interesting to give finer bounds for restricted classes of functions. This problem has indeed been studied for $\dnf$'s with a bounded number of terms. It is known that a $\dnf$ with $k$ terms has at most $2^k - 1$ prime implicants (see e.g. \cite{sst}). 

In this note we consider the same problem for the class of $k$-${\cnf}$ functions. Understanding the structure of the set of satisfying assignments of $k$-$\cnf$ formulas has been a crucial subject in computational complexity, in particular in developing $k$-SAT algorithms and bounded depth circuit lower bounds. Notable examples are the characterization of satisfying assignments of a 2-$\cnf$ which yields a polynomial time algorithm for 2-SAT (see e.g. \cite{apt}), and the Satisfiability Coding Lemma of \cite{ppz} which bounds the number of isolated satisfying assignments of $k$-${\cnf}$s (these are the assignments such that if we flip the value of any single variable, the formula is not satisfied anymore). This was later improved in \cite{ppsz} to obtain the best known $k$-SAT algorithm and depth-3 circuit lower bounds for an explicit function. 


To describe our main results we need a few definitions which follow shortly. 
 
A {\it restriction} on a set $X$ of variables is a mapping $\rho: X \rightarrow \{*, 0, 1\}$. We call a variable {\it free} if it is assigned $*$, and we call it {\it fixed} otherwise. We say that a restriction is {\it partial} if it leaves at least one variable free. For a function $f$ over a set $X$ of variables and a restriction $\rho$, we define $f_{\rho}$ to be the subfunction obtained after setting values to the fixed variables according to $\rho$. An {\it implicant} of $f$ is a restriction $\rho$ such that $f_{\rho}$ is the constant 1 function. We define a {\it prime implicant} of $f$ to be an implicant $\rho$ of $f$ such that unspecifying any fixed variable does not yield the constant 1 function. A {\it partial prime implicant} is a prime implicant that leaves at least one variable free. To see that the concept of prime implicant generalizes that of isolated satisfying assignments, note that any isolated solution is in fact a prime implicant. Furthermore if $\rho$ is a partial prime implicant, it is easy to see that if we remove all the variables in $\rho^{-1}(*)$ from the formula, then $\rho$ restricted to $X \setminus \rho^{-1}(*)$ is in fact an isolated solution of this derived formula. 


The following lemma due to Paturi, Pudl{\'a}k and Zane gives a bound on the number of isolated solutions of a $k$-$\cnf$.

\begin{lm}[The Satisfiability Coding Lemma \cite{ppz}]\label{ppzl}
Any $k$-$\cnf$ on $n$ variables has at most $2^{(1 - \frac{1}{k})n}$ isolated satisfying assignments.
\end{lm}

In an attempt to extend this result we define $m(n, k)$ to be the maximum number of prime implicants over $k$-$\cnf$ formulas on $n$ variables. It is a natural question to give sharp bounds for $m(n, k)$. A similar problem was studied by Miltersen, Radhakrishnan and Wegener \cite{mrw} which asks for the smallest size of a $\dnf$ equivalent to a given $k$-$\cnf$. We first give a lower bound for $m(n, k)$.

\begin{prp}
There exists $k_0$ such that for all $n \ge k \ge k_0$, $m(n, k) \ge 3^{(1 - O(\frac{\log k}{k}))n}$.
\end{prp}

\begin{pf}
We follow the construction of Chandra and Markowsky \cite{chand}. We divide the set of $n$ variables in $n/k$ parts, each of size $k$. On each of these parts, we represent the Chandra-Markowsky function as a $k$-$\cnf$, that is the disjunction of all conjunctions of $2k/3$ variables, exactly $k/3$ of which are negated. We can do this since each such function depends only on $k$ variables. Formula $F$ would then be obtained by conjuncting all these functions together. In \cite{chand} it was shown that each block has at least $\Omega(3^k/k)$ prime implicants. It is easy to see that prime implicants of $F$ are obtained by concatenating prime implicants of the blocks. Therefore the total number of prime implicants is at least $\Omega((3^k/k)^{n/k}) = 3^{(1 - O(\log k / k))n}$. \qed
\end{pf}

For $k=2$ we manage to give almost tight bounds. Note that in this case the above bound is not applicable as it is only valid as long as $k$ is large.

\begin{thm}\label{main}
$3^{\frac{n}{3}} \le m(n, 2) \le (1+o(1))3^{\frac{n}{3}}$.
\end{thm}


\section{Proof of Theorem \ref{main}}

We first prove the lower bound. Let $n = 3m$ and consider the following formula on variable set $\{x_1, \ldots, x_m, y_1, \ldots, y_m, z_1, \ldots, z_m\}$ suggested to us by Dominik Scheder: $$T(x,y,z) = \bigwedge_{i = 1}^m (x_i \vee y_i) \wedge (y_i \vee z_i) \wedge (x_i \vee z_i).$$ It is easy to see that every prime implicant of $T$ and every $1 \le i \le m$ must set exactly two variables among $x_i$, $y_i$ and $z_i$ to 1. Therefore $T(x, y, z)$ has $3^{\frac{n}{3}}$ prime implicants.

We now move on to the upper bound. Let $F$ be any Boolean function on $\{x_1, \ldots, x_n\}$ and let $\rho$ be a prime implicant that fixes all the variables. We claim that $\rho$ is an {\it isolated} satisfying assignment for $F$, that is if we change the value of any single one of the variables, the formula evaluates to 0. To see this note that if changing the value of some variable $x_i$ still satisfies $F$, we can simply unspecify $x_i$ and get a smaller restriction which yields the constant 1 function, contradicting the minimality of $\rho$. When $F$ is a 2-$\cnf$ we can apply Lemma \ref{ppzl} and bound the number of such prime implicants by $2^{\frac{n}{2}}$.

It thus remains only to bound the number of partial prime implicants. Assume without loss of generality that $F$ contains no clauses with only one literal, since the value of such literal is forced. We need some terminology which we borrow from \cite{apt}. We define the {\it implication digraph} of $F$ which we denote by $D(F)$ as follows. The vertex set consists of all literals $x_1, \overline{x}_1, \ldots, x_n, \overline{x}_n$. For every clause $x \vee y$ in $F$ we put two directed edge $\overline{x} \rightarrow y$ and $\overline{y} \rightarrow x$ in $D(F)$. We say that a literal $u$ {\it implies} a literal $v$, if there exists a directed path from $u$ to $v$ in $D(F)$. From here on we will assume without loss of generality that $D(F)$ is loopless. This is because a loop in $D(F)$ corresponds to a clause of the form $u \vee \overline{u}$ which is always true. It is easy to see that one can characterize the set of satisfying assignments of 2-$\cnf$s in terms of their implication digraphs. 

\begin{prp}[\cite{apt}]\label{impd}
An assignment $\alpha$ satisfies a 2-$\cnf$ $F$ if and only if there is no edge in $D(F)$ going out of the set of true literals.
\end{prp} 

We now give a similar characterization of partial prime implicants. For a restriction $\rho$ we partition the set of literals into three sets $A_{\rho}$, $B_{\rho}$ and $C_{\rho}$ containing false literals, true literals, and those that are free, respectively. For any set $S$ of vertices let $N^{-}(S)$ and $N^{+}(S)$ be the set of literals not in $S$ implying some literal in $S$ and the set of literals not in $S$ implied by some literal in $S$, respectively. We first make the following observation.

\begin{prp}\label{size}
Let $F$ be a 2-$\cnf$ and let $S$ be the set of all literals that appear in some directed cycle in $D(F)$. Then for any implicant $\rho$ of $F$, we have $S \cap C_{\rho} = \emptyset$.
\end{prp}

\begin{pf}
Fix any directed cycle $T$ in $D(F)$. In any satisfying assignment of $F$, all literals in $T$ should be assigned the same value, since otherwise there would be a path connecting a true literal to a false one. Similarly, in an implicant of $F$, if a literal in $T$ is free then in fact all literals in $T$ must be free, since fixing the value of any such literal forces the value of all other literals in $T$. But this means that those clauses that contain variables only in $T$ are completely untouched by $\rho$ and hence not satisfied. This contradicts the assumption that $\rho$ is an implicant. \qed
\end{pf}

\begin{prp}\label{char-implicant}
Let $F$ be a 2-$\cnf$. A restriction $\rho$ is an implicant of $F$ if and only if
\begin{enumerate}
\item there is no edge into $A_{\rho}$
\item there is no edge out of $B_{\rho}$
\item $C_{\rho}$ is an independent set.
\end{enumerate}
\end{prp}

\begin{pf}
$\Rightarrow$: Assume that $\rho$ is a partial implicant. Assume for a contradiction that there is an edge going into $A_{\rho}$, namely $u \rightarrow v$. By construction of $D(F)$ we have $\overline{u} \vee v$ as a clause in $F$. Assume $u \in B_{\rho}$. But then $\rho$ leaves this clause unsatisfied which is a contradiction. If $u \in C_{\rho}$, since it is left free by $\rho$ we can set it to 0 and we will get a $1 \rightarrow 0$ edge, and hence a contradiction with the same argument. If there is an edge out of $B_{\rho}$, by a symmetric argument we can get a $1 \rightarrow 0$ edge and reach a contradiction. Furthermore, $C_{\rho}$ is an independent set, since otherwise there would be a clause which is left completely untouched by $\rho$. 

$\Leftarrow$: We need to show that $\rho$ satisfies the formula. Notice that all the clauses are hit by $\rho$, since $C_{\rho}$ is an independent set. To see that the formula is indeed satisfied, consider an arbitrary clause $u \vee v$. If both $u$ and $v$ are fixed by $\rho$, since there is no edge from a true literal to a false literal, the clause should evaluate to true. If $u$ is fixed but $v$ is left free, $u$ has to be set to true, since otherwise there will be an edge from $\overline{u}$ to $v$ contradicting the fact that there is no edge going out of $B_{\rho}$. If $u$ is free and $v$ is fixed we can use a similar argument.  \qed
\end{pf}

\begin{prp}\label{char}
Let $F$ be a 2-$\cnf$ such that $D(F)$ is acyclic. A restriction $\rho$ is a partial prime implicant of $F$ if and only if 
\begin{enumerate}
\item there is no edge into $A_{\rho}$
\item there is no edge out of $B_{\rho}$
\item $C_{\rho}$ is a non-empty independent set such that $A_{\rho} = N^-(C_{\rho})$ and $B_{\rho} = N^+(C_{\rho})$.

\end{enumerate}
\end{prp}

\begin{pf}
$\Rightarrow$: Since $\rho$ is an implicant, item (1) and (2) and that $C_{\rho}$ is an independent set follow from Proposition \ref{char-implicant}. Since $\rho$ is a partial prime implicant we have $C_{\rho} \ne \emptyset$. Let $u_1 \in A_{\rho}$ be a literal that does not imply any literal in $C_{\rho}$. There must exist another literal $u_2 \in A_{\rho}$ such that $u_1 \rightarrow u_2$, since otherwise we can make a restriction $\rho'$ by unfixing $u_1$. It is easy to see that $\rho'$ satisfies $F$ which contradicts the minimality of $\rho$. But now $u_2$ cannot imply any literal $C_{\rho}$ since otherwise there would be a path from $u_1$ to $u_2$ and then to a node in $C_{\rho}$, contradicting the assumption. Using this argument we can get an infinite sequence $u_1 \rightarrow u_2 \rightarrow \ldots$ of nodes in $A_{\rho}$. But since $A_{\rho}$ is finite, there must exist a cycle on this sequence and hence a contradiction. With a similar argument we can prove $B_{\rho} = N^+(C_{\rho})$. 

$\Leftarrow$: By Proposition \ref{char-implicant} and the fact that $C_{\rho}$ is non-empty, $\rho$ is a partial implicant. To see that it is a prime implicant, note that we cannot unfix a literal and still satisfy the formula, since any such literal implies or is implied by some free literal.  \qed
\end{pf}

\begin{lm}\label{rest}
Let $F$ be a 2-$\cnf$ and let $\rho$ be any partial prime implicant of it. Let $F'$ be the 2-$\cnf$ obtained by considering those clauses of $F$ that contain only literals that do not belong to any directed cycle in $D(F)$. Assume further that the corresponding variables to all such literals do appear in $F'$. Then $\rho$ restricted to these literals is a prime implicant of $F'$. 
\end{lm}

\begin{pf}
Let $T$ be the set of literals of $F$ that do not appear in any directed cycle. Note that $F'$ is already satisfied by $\rho$, since $\rho$ satisfies all clauses in $F$. We then only need to show that if we unfix any literal in $T$, some clause in $F'$ will be left unsatisfied. By Proposition \ref{size} we have $C_{\rho} \subseteq T$. Let $u \in T$ be a literal that has been assigned a value by $\rho$. By Proposition \ref{char}, we have either $u \in N^-(C_{\rho})$ or $u \in N^+(C_{\rho})$. Without loss of generality assume that $u \in N^-(C_{\rho})$. This implies that there exists a literal $v \in C_{\rho}$ such that $u$ implies $v$. If $u$ directly connects to $v$, this corresponds to having a clause $\overline{u} \vee v$ in $F$ and hence in $F'$. But making $u$ free will make this clause unsatisfied. If $u$ is not directly connected to $v$, there exists some $u' \in T$ such that there is an edge from $u$ to $u'$ and $u'$ implies $v$. This is so since otherwise $u$ would not appear in $F'$. The $u \rightarrow u'$ edge corresponds to a clause $\overline{u} \vee u'$ in $F$. Applying Proposition \ref{char-implicant}, we know that $u'$ is set to 0 under $\rho$. If we unfix $u$, the clause $\overline{u} \vee u'$ is not satisfied anymore. This finishes the proof. \qed
\end{pf}


\begin{lm}\label{cnt}
Let $F$ be a 2-CNF on $n$ variables with no directed cycle in $D(F)$. Then the number of partial prime implicants of $F$ is upper bounded by $3^{n/3}$.
\end{lm}

\begin{pf}
From $F$ we construct a graph $G(F)$ on vertices $\{x_1, \ldots, x_n\}$ and we include an edge $(x_i, x_j)$ if and only if there are literals $u$ and $v$ on $x_i$ and $x_j$, respectively, such that either of $u$ or $v$ implies the other. We give an injective mapping from the set of partial prime implicants of $F$ to the set of maximal independent sets of $G(F)$. Let $\rho$ be a partial prime implicant of $F$. Let $S$ be the set of distinct variables appearing in $C_{\rho}$. We map $\rho$ to $S$ and claim that this mapping satisfies the required properties. We first show that $S$ forms a maximal independent set in $G(F)$. If there is an edge $(x, y)$ in $S$, there is a path between literals on $x$ and $y$ in $D(F)$. By Proposition \ref{char}, $C_{\rho}$ is an independent set. Thus this path has to go out of $C_{\rho}$ and hence to $N^+(C_{\rho})$. But since by Proposition \ref{char} there is no edge going out of $N^+(C_{\rho})$, we cannot continue this path, a contradiction. To prove maximality, assume that there exists a vertex $x \not \in S$ that is not adjacent to any vertex in $S$. By construction this implies that there is no path in $D(F)$ between neither of $x$ nor $\overline{x}$ and $C_{\rho}$. But this cannot happen since by Proposition \ref{char} the set of literals is partitioned into $N^{-}(C_{\rho}) \cup N^{+}(C_{\rho}) \cup C_{\rho}$. To show that this mapping is injective, assume that there are two distinct partial prime implicants $\rho$ and $\rho'$ that are mapped to the same set $S$. This means that $C_{\rho} = C_{\rho'}$. But by Proposition \ref{char} we have $A_{\rho} = N^-(C_{\rho}) = N^-(C_{\rho'}) = A_{\rho'}$ and $B_{\rho} = N^+(C_{\rho}) = N^+(C_{\rho'}) = B_{\rho'}$. This implies that $\rho = \rho'$.

It thus remains to bound the number of maximal independent sets in $G(F)$, which is just a simple undirected graph on $n$ vertices. This is a very well-known problem for which sharp bounds are known.

\begin{thm}[Moon, Moser \cite{mm}]
\label{numind}
In any graph on $n$ vertices, there are at most $3^{\frac{n}{3}}$ maximal independent sets.
\end{thm}

Applying Moon-Moser theorem we obtain that the number partial prime implicants of $F$ is upper bounded by $3^{n/3}$. \qed
\end{pf}

Let $t$ be the number of distinct variables that belong to some directed cycle in $D(F)$. Under any prime implicant these variables are fixed. Since all literals in a cycle should be assigned the same value, and there are at least 2 distinct variables in each, there are at most $2^{t/2}$ ways to fix them. Now we consider the rest of the literals. We notice that after fixing the first $t$ variables,  any literal that does not appear in any cycle and is connected only to some literals that appear in a cycle, is uniquely set in any prime implicant. This is because if such literal implies a literal which is set to 0 then it has to be set to 0, if it is implied by a literal set to 1, it has to be set to 1, and otherwise it has to be left free. For the rest of the variables we can now apply Lemma \ref{rest} and Lemma \ref{cnt} and bound the number of partial prime implicants by $2^{t/2} \cdot 3^{(n - t)/3} \le 3^{n/3}$.

The set of prime implicants consists of those that are partial and those that fix all the variables. Therefore the total of number of prime implicants is bounded by $3^{\frac{n}{3}} + 2^{\frac{n}{2}}$.

\section{Concluding Remarks}
We gave an essentially sharp bound on the number of prime implicants of 2-$\cnf$ formulas. It remains an interesting question to obtain a similar bound for $k \ge 3$. To do so we need to develop a technique that can treat isolated solutions and prime implicants in general as the same objects.

\section{Acknowledgements}
I am grateful to Dominik Scheder and Stefan Schneider for useful discussions and to Gy{\"o}rgy Tur{\'a}n for correspondence. I would also like to thank Kristoffer Arnsfelt Hansen for pointing out an error in an earlier draft and to anonymous reviewers for their helpful comments.

\bibliographystyle{elsarticle-num}
\bibliography{ref}

\end{document}